\documentclass[oldversion]{aa} 
\usepackage{graphicx}
\usepackage{txfonts}
\usepackage{natbib}
\bibpunct{(}{)}{;}{a}{}{,}

\pdfoutput=1

\begin{document}
 
\title{A high resolution spectral atlas of brown dwarfs\thanks{Based on observations 
    collected at the European Southern Observatory, Paranal, Chile, 077.C-0449}}

\author{A.~Reiners
  \inst{1}\fnmsep\thanks{Emmy Noether Fellow}
  \and
  D.~Homeier\inst{1}
  \and
  P.~H.~Hauschildt\inst{2}
  \and
  F.~Allard\inst{3}
}

\offprints{A. Reiners}

\institute{
  Universit\"at G\"ottingen, Institut f\"ur Astrophysik, Friedrich-Hund-Platz 1, D-37077 G\"ottingen, Germany\\
  \email{Ansgar.Reiners@phys.uni-goettingen.de, derek@astro.physik.uni-goettingen.de}
  \and
  Hamburger Sternwarte, Gojenbergsweg 112, D-21029 Hamburg, Germany
  \and
  Centre de Recherche Astrophysique de Lyon, UMR 5574: CNRS, Universit\'e de Lyon,
                   \'Ecole Normale Sup\'erieure de Lyon, 46 all\'ee d'Italie, F-69364 Lyon Cedex 07, France\\
}

\date{Received May 5, 2007 / Accepted July 13, 2007}


\abstract {We present a UVES/VLT high resolution atlas of three L
  dwarfs and one T dwarf system, spectral classes at which most of the
  objects are brown dwarfs. Our atlas covers the optical region from
  H$\alpha$ up to the near infrared at 1\,$\mu$m. We present spectral
  details of ultra-cool atmospheres at very high resolution ($R \sim
  33\,000$) and compare the spectra to model calculations.  Our
  comparison shows that molecular features from VO and CaH, and atomic
  features from Cs and Rb are reasonably well fit by current models.
  On the other hand, features due to TiO, CrH, and water, and atomic
  Na and K reveal large discrepancies between model calculations and
  our observations. }


\maketitle
%

\section{Introduction}

The analysis of brown dwarfs has seen tremendous progress since their
first detection more than ten years ago. Since then, the stellar
spectral sequence was augmented by spectral classes L and T at its
cool end. L and T objects must certainly be cooler than those
classified as spectral type M, but the reasons for the spectral
evolution through the L- and T-classes turned out to be more complex
than in the hotter objects \citep[see e.g., ][]{Kirkpatrick05}. The
formation of molecules sets in at temperatures typical for spectral
classes K and M, and their opacities begin to cause strong deviations
from blackbody spectra. Furthermore, the condensation of dust clouds
becomes important at the transition from M to L and dominates the
appearance of spectral class L objects. At the L to T transition, the
atmospheric effects of the dust grains weaken, but this transition is
still poorly understood \citep{Burrows06}.

Objects of spectral class T are always brown dwarfs according to
standard evolutionary models \citep{BDReview2000}.  Depending on age,
however, objects of spectral classes M or L may be either brown dwarfs
or hydrogen-burning stars. Young brown dwarfs can be found in the same
temperature range as old low-mass stars, hence it is not possible to
determine brown dwarf status from the spectral type (taken as proxy
for effective temperature) alone \citep{Basri00ARAA, Chabrier05}.

As brown dwarfs are defined by mass, there is no obvious atmospheric
feature that determines brown dwarf status by just looking at an
object's spectrum.  While the Li-test \citep[e.g.,][]{Basri00ARAA} is
a useful tool in young objects, it is not suitable for the majority of
old (field) objects. Age can as well be determined in members of
stellar clusters, although by today this is only useful for young
brown dwarfs because they are bright enough to be observed at the
large distance of the clusters. A direct determination of brown dwarf
status is possible for members of binary systems through the
measurement of their mass, but that is not possible for most field
objects. The obvious way to determine the mass of an ultra-cool object
is through its gravity \citep[e.g.,][]{Mohanty04}. The mass can
directly be calculated once gravity and radius are known, and the
latter can be obtained from the luminosity, a parameter that is
relatively well constrained.  \cite{Burgasser06} have shown a method
to determine the gravity in T dwarfs, and they calculate masses for 13
T dwarfs.  These authors determine gravity from color ratios in low
resolution spectra that they calibrated from model calculations. They
claim rather low uncertainties for their results, but one has to keep
in mind that this method is strongly dependent on the match between
calibration models and observed spectra \citep[see also discussion of
uncertainties in][]{Burgasser07}. The models of ultra-cool
atmospheres, however, are still evolving, and ``the extraction of
physical quantities such as $T_{\rm eff}$, gravity, and metallicity
from the growing library of well-calibrated spectra, although possible
in a crude sense, remains imprecise.''  \citep{Burrows06}

The rapid progress in the investigation of (ultra-)cool atmospheres is
due to the advent of 10m class telescopes that allow spectroscopy in
these faint targets. The redward shift of the spectral energy
distribution (SED) in cool stars moves the peak of the SED out of the
range of ``classical'' optical spectrographs into the IR domain.
Furthermore, the drop in temperature is connected to a loss in
luminosity with $T^4$, and both effects together render it very
difficult to obtain high quality optical spectroscopic data for the
coolest stars and brown dwarfs.

\cite{Kirkpatrick91} have shown low resolution red and near-IR spectra
in mid-K and M dwarfs, and they have also identified a great many
atomic lines and molecular bands. Optical low resolution spectra of T
dwarfs together with identifications of the main absorption features
were provided by \cite{Burgasser03}. For the IR wavelength range,
\cite{McLean03} and \cite{Cushing05} published compilations of low
resolution spectra in M, L and T dwarfs, and \cite{McLean07} also
showed high resolution IR spectra for these spectral classes.

High resolution spectra of ultra-cool stars and brown dwarfs are
difficult to obtain. This is a pity because it is the pressure
dependent wings of the saturated atomic absorption lines
(predominantly of the alkali elements) that are most sensitive to
gravity. While the strong alkali lines depend on effects like
temperature, metallicity, rotation, etc. as well, these parameters can
be determined from molecular bands, which can resolve the degeneracy
between gravity and other parameters. In high resolution spectra,
several features are available to constrain atmospheric parameters
simultaneously. Thus, the determination of gravity from the pressure
broadened wings can be expected to be much more accurate than
comparing color ratios from low resolution spectra.  High resolution
optical spectra of M dwarfs were shown by \cite{Tinney98}, but no high
resolution data showing the full optical spectrum in L dwarfs or
cooler were made available. Some individual features in high
resolution L dwarf spectra were investigated by \cite{Basri00,
  Schweitzer01, Schweitzer02}.

In this paper, we present an atlas of high resolution, high signal to
noise spectra in the red and near-IR spectral range for three L and an
early T dwarf system ($2200 \la T_{\rm eff} \la 1200$\,K). We also
compare a set of the most recent model calculations to the data.  For
the full wavelength region, we show the spectral features of
ultra-cool L and early T dwarfs at high resolution, and we make a
first attempt to identify the shortcomings and successes of our
current model calculations.

\section{Data}
\subsection{Objects and format of the atlas}

We present a high resolution spectral atlas of four very low mass
objects of spectral types between L0 and T1. Name, spectral type,
$J$-magnitude, and exposure times of the objects are given in
Table\,\ref{tab:observations}. The signal to noise ratio (SNR) of the
spectra varies over the wavelength region according to the object's
spectral energy distribution and detector efficiency, it is well over
20 in most of the region in the L dwarfs, so that the dense molecular
absorption features are well discernible from the noise.

\begin{table}[h]
  \begin{minipage}[t]{\columnwidth}
    \centering
    \caption{\label{tab:observations}Objects observed for this atlas and temperatures adopted for the models.}
    \renewcommand{\footnoterule}{}  
    \begin{tabular}{ccccc}
      \hline
      \hline
      \noalign{\smallskip}
      Object & SpType\footnote{optical spectral type from \texttt{DwarfArchives.org}} & $J$\footnote{2 MASS system}& exp. time& $T_{\rm{model}}$\footnote{see Section~\ref{sect:models}}\\
      &&[mag]&[s]&[K]\\
      \noalign{\smallskip}
      \hline
      \noalign{\smallskip}
      2MASS 1731$+$2721 & L0 & 12.09 & 2400 & 2200\\
      2MASS 1155$-$3727 & L2 & 12.81 & 5200 & 2000\\
      2MASS 1507$-$1627 & L5 & 12.83 & 5200 & 1700\\
      2MASS 2204$-$5646 & T1+T6 & 11.91 & 3000 & 1280\\
      ($\epsilon$ Indi\,B)\\
      \noalign{\smallskip}
      HD 114772 & B9V & $V$=5.89 & 25 &\\
      \noalign{\smallskip}
      \hline
    \end{tabular}
  \end{minipage}
\end{table}

\begin{figure*}
  \resizebox{.97\hsize}{!}{\includegraphics[angle=90]{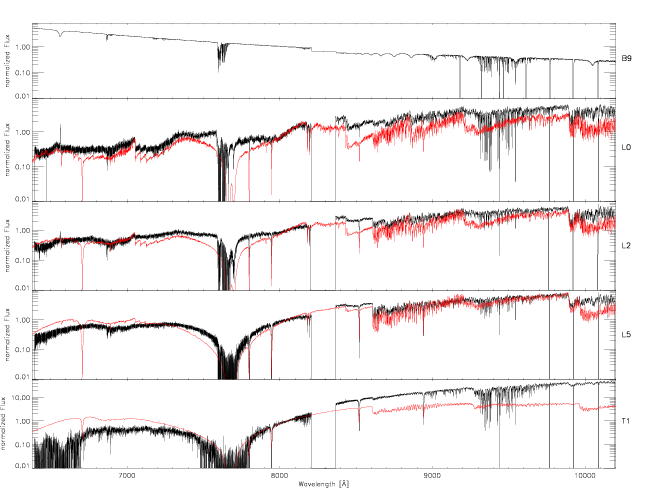}}
  \caption{Full spectra (black) and models (red) of 2M1731
    (L0), 2M1155 (L2), 2M1507 (L5), and 2M2204 (T1). For comparison, a
    telluric reference star (spectral type B9V) is shown.}
  \label{fig:atlas_full}
\end{figure*}

\begin{figure*}
  \resizebox{.97\hsize}{!}{\includegraphics[angle=90]{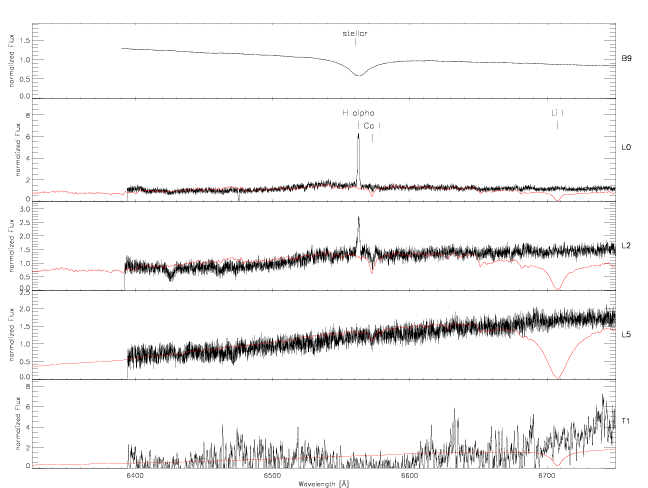}}
  \caption{Spectral atlas (black) and models (red) of 2M1731 (L0), 2M1155 (L2),
    2M1507 (L5), and 2M2204 (T1). For comparison, a telluric reference
    star (spectral type B9V) is shown.}
  \label{fig:atlas1}
\end{figure*}

\begin{figure*}
  \resizebox{.97\hsize}{!}{\includegraphics[angle=90]{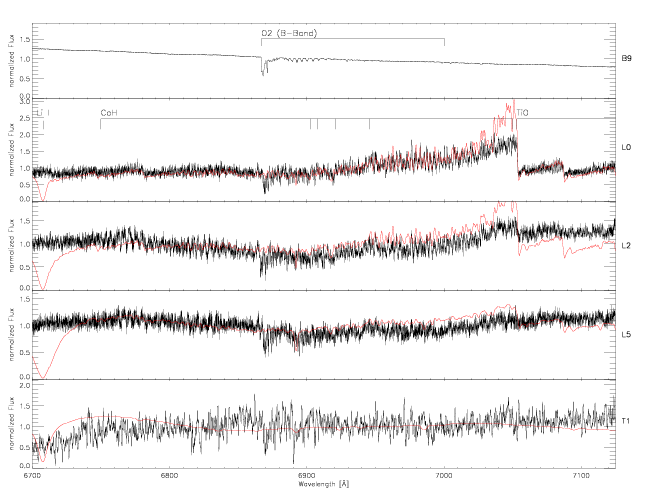}}
\caption{Spectral atlas (black) and models (red) of 2M1731 (L0), 2M1155 (L2), 2M1507 (L5), and 2M2204 (T1) (continued).}
  \label{fig:atlas2}
\end{figure*}

\begin{figure*}
  \resizebox{.97\hsize}{!}{\includegraphics[angle=90]{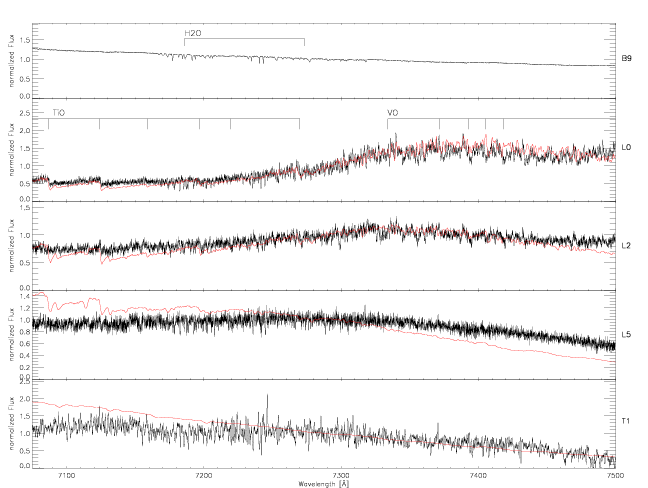}}
\caption{Spectral atlas (black) and models (red) of 2M1731 (L0), 2M1155 (L2), 2M1507 (L5), and 2M2204 (T1) (continued).}
  \label{fig:atlas3}
\end{figure*}

\begin{figure*}
  \resizebox{.97\hsize}{!}{\includegraphics[angle=90]{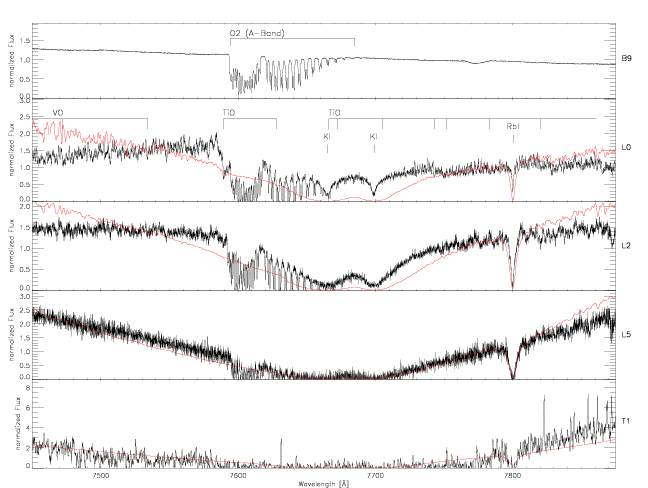}}
\caption{Spectral atlas (black) and models (red) of 2M1731 (L0), 2M1155 (L2), 2M1507 (L5), and 2M2204 (T1) (continued).}
  \label{fig:atlas4}
\end{figure*}

\begin{figure*}
  \resizebox{.97\hsize}{!}{\includegraphics[angle=90]{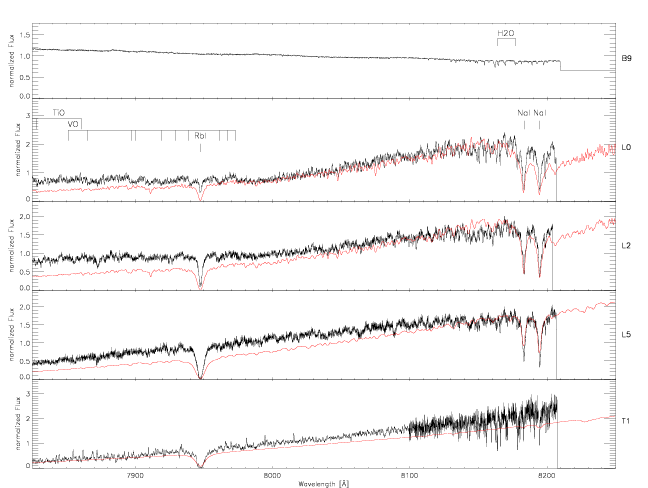}}
\caption{Spectral atlas (black) and models (red) of 2M1731 (L0), 2M1155 (L2), 2M1507 (L5), and 2M2204 (T1) (continued).}
  \label{fig:atlas5}
\end{figure*}

\begin{figure*}
  \resizebox{.97\hsize}{!}{\includegraphics[angle=90]{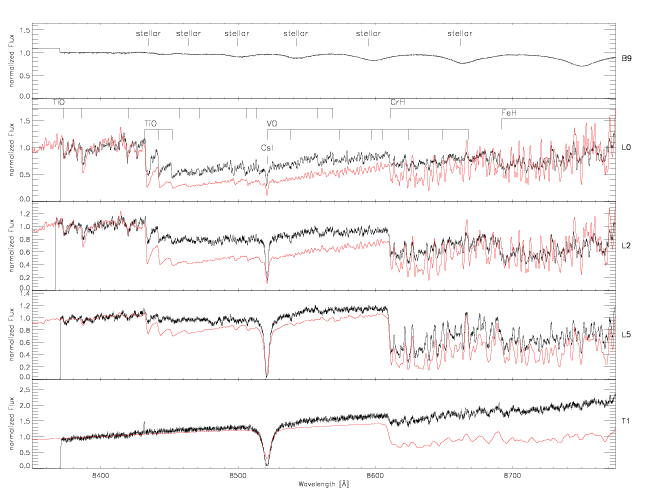}}
\caption{Spectral atlas (black) and models (red) of 2M1731 (L0), 2M1155 (L2), 2M1507 (L5), and 2M2204 (T1) (continued).}
  \label{fig:atlas6}
\end{figure*}

\begin{figure*}
  \resizebox{.97\hsize}{!}{\includegraphics[angle=90]{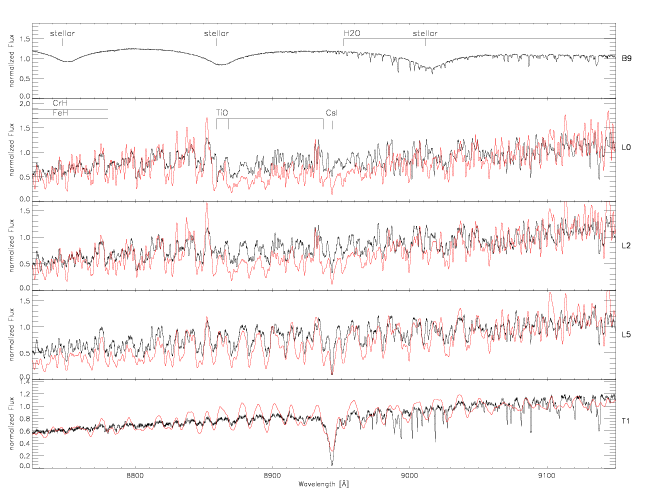}}
\caption{Spectral atlas (black) and models (red) of 2M1731 (L0), 2M1155 (L2), 2M1507 (L5), and 2M2204 (T1) (continued).}
  \label{fig:atlas7}
\end{figure*}

\begin{figure*}
  \resizebox{.97\hsize}{!}{\includegraphics[angle=90]{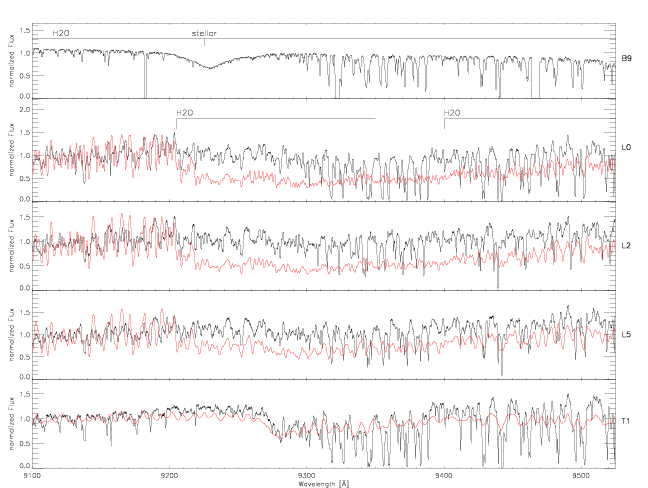}}
\caption{Spectral atlas (black) and models (red) of 2M1731 (L0), 2M1155 (L2), 2M1507 (L5), and 2M2204 (T1) (continued).}
  \label{fig:atlas8}
\end{figure*}

\begin{figure*}
  \resizebox{.97\hsize}{!}{\includegraphics[angle=90]{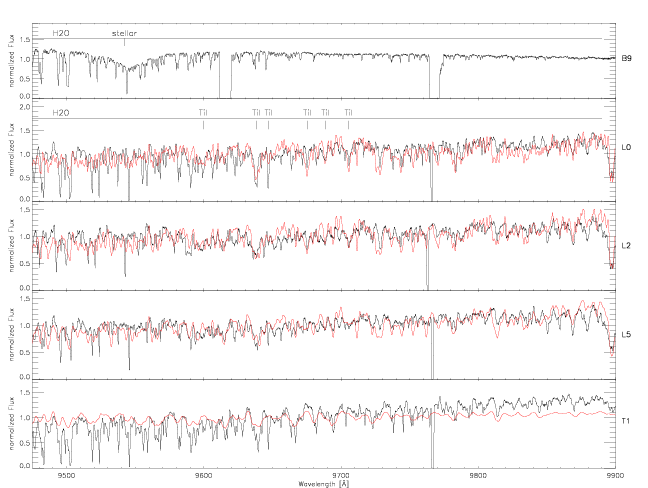}}
\caption{Spectral atlas (black) and models (red) of 2M1731 (L0), 2M1155 (L2), 2M1507 (L5), and 2M2204 (T1) (continued).}
  \label{fig:atlas9}
\end{figure*}

\begin{figure*}
  \resizebox{.97\hsize}{!}{\includegraphics[angle=90]{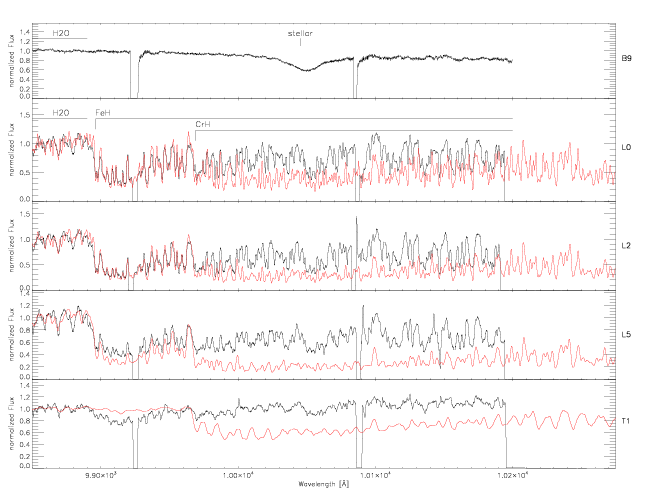}}
\caption{Spectral atlas (black) and models (red) of 2M1731 (L0), 2M1155 (L2), 2M1507 (L5), and 2M2204 (T1) (continued).}
  \label{fig:atlas10}
\end{figure*}

In Fig.\,\ref{fig:atlas_full}, we show the full spectra of all four
targets. In the top panel, we also provide the spectrum of a rapidly
rotating B9V star (also listed in Table\,\ref{tab:observations}) that
we observed with the same setup to identify telluric lines due to
absorption of Earth's atmosphere. In Figs.\,\ref{fig:atlas1} to
\ref{fig:atlas10} we provide the detailed spectral atlas with
identifications of the most prominent absorption features. In all
plots, synthetic spectra chosen to match the object's effective
temperature are overplotted. We discuss the models in
\S\,\ref{sect:models}.

\subsection{Observations}

Data were obtained with the UVES spectrograph at ESO/VLT in service
mode during April and May, 2006. UVES was operated in dichroic mode
using both the blue and the red arm. Because of the red colour of L
dwarf spectra, the blue instrument arm contains very little flux if
any, but it covers several hydrogen lines that are interesting during
flares. For this spectral atlas, we only consider the red arm that was
configured in a non-standard setting centered at 830\,nm. This setting
covers the wavelength region 6400 -- 10\,200\,\AA, it contains
H$\alpha$ and the FeH Wing-Ford band that is very useful for spectral
analysis in late-type dwarfs \citep{Reiners06a}. Between the two CCDs
of the red arm, the spectra have a gap from 8200\,\AA\ to 8370\,\AA.
The Na doublet at 8190\,\AA\ lies just blueward of the gap. The
spectra were taken at a slit width of 1\farcs 2 yielding a resolving
power of $R \sim 33\,000$. Data were reduced using the MIDAS-based
ESO-pipeline for UVES data. Bias-subtraction, division by a
flatfield-lamp spectrum, and wavelength calibration follow standard
routines. We did not remove telluric lines for our spectral atlas, but
we show a spectrum of a telluric standard star for comparison.

\section{Models}
\label{sect:models}
Theoretical models have been calculated using the general-purpose
stellar atmosphere code \texttt{PHOENIX} version 15.2. Details of the
numerical methods are given in \citet{jcam}. In this work we use a
setup of the microphysics that gives the currently best fits to
observed spectra of M, L, and T dwarfs for the low $T_{\rm eff}$
regime and that also updates the microphysics used in the GAIA model
grid \cite[]{2005A&A...442..281K,2006A&A...452.1021K}.  The water
lines are taken from the calculations of \citet{bt2}; this list gives
the best overall fit to the water bands over a wide temperature range.
TiO lines are taken from \cite{ames-tio} for similar reasons.  The
treatment of opacities and the equation of state (EOS) is similar to
the one described in detail in \cite{LimDust}, with extensions and
modifications described below (HITRAN
database\footnote{\texttt{http://www.cfa.harvard.edu/HITRAN/}}; Plez,
Bernath, priv.comm.).

One of the most important recent improvements of cool stellar
atmosphere models is the availability of new atomic line profile data
based on accurate inter-atomic potentials.  The calculations presented
here include detailed and depth dependent line profiles for each of
the D1 and D2 transitions of the Alkali resonance doublets (Li, Na, K,
Rb, Cs) for perturbation by both H$_2$ and He, as described in
\citet{Alkalis03,alkalisLi,Allard06,Cores07,Allard07a}.  We also
include damping of the Na and K lines by H\textsc{i} using the
broadening widths of \citet{1991SoPh..131....1A,2000JPhB...33.1805L}
and \citet{2001JPhB...34.4785B}, though for L dwarf temperatures
atomic hydrogen only accounts for less than 10\% of the total
broadening \citep{Johnas07}.

These profiles give a much improved representation of the
details of the line shapes in the optical spectra of M, L and T
dwarfs, and become especially 
important in situations where line blanketing and broadening are
crucial for the model structure calculations and for the computation
of the synthetic spectra, i.\,e.\ most notably in T dwarfs, where dust
as a source of continuum opacity becomes less important. 

The EOS is an updated version of the one used in \citet{LimDust}, we
include about 1000 species (atoms, ions and molecules). The EOS
calculations themselves follow the method discussed in \citet{MDpap}.
For effective temperatures $T_{\rm eff} < 2500$ K, the condensation of
dust particles has to be considered in the EOS. In our models we allow
for the formation (and dissolution) of a variety of grain species. For
details of the EOS and the opacity treatment of condensates see
\citet{LimDust}.

It has long been evident that the 'Dusty' limit described in
\citet{LimDust}, which assumes the presence of condensates everywhere
where they are thermodynamically stable, becomes increasingly
inaccurate in the case of late L and especially T dwarfs.  A more
realistic treatment needs to include the gravitational settling,
sedimentation or rainout, of condensates, which results in a cloud
layer retreating to larger optical depths as $T_{\rm eff}$ decreases
\citep{settlIAU211}.  While these processes become very noticeable at
$T_{\rm eff}$ below 2000\,K, the settling of dust in high altitudes
and the associated depletion of condensable elements can affect the
spectra already much earlier for features that form at very low
optical depths, such as strong molecular bands and the cores of the
Na\,I and K\,I resonance lines.  \citet{LoddersAU2} have shown that
the sequential removal of refractory species in order of decreasing
condensation temperature -- the ``rainout'' scenario -- leads to
different chemical composition from the full chemical equilibrium (CE)
treatment, because in the former case high temperature condensates
would remove their constituent elements from the gas phase in deep
layers of the atmosphere, thus cutting off all higher layers from the
supply of these elements.  The formation of lower condensation
temperature species, that would form in a fully mixed gas in CE, could
thus be inhibited.  One such case potentially relevant for brown dwarf
atmospheres is the formation of anorthite (CaAl$_2$Si$_2$O$_8$), which
would allow alkali metals to condense into feldspars (e.\,g.\ albite,
NaAlSi$_3$O$_8$, orthoclase, KAlSi$_3$O$_8$) at $T<1200$\,K.  However,
other calcium silicates can form at much hotter temperatures, as soon
as at the 1700\,K level. Thus, in a fully rained out atmosphere, it
will remove all Ca from the gas phase correspondingly deeper in the
atmosphere, and from all layers above that level.  Anorthite would
thus not get a chance to form. Therefore, in such a chemistry the
first alkali-bearing condensates to appear are the alkali halides at
$\sim$\,1000\,K. The alkali metals could then only be depleted at
higher and cooler layers than in the CE case, where feldspars could
form around $\sim$\,1200\,K.

In a real (sub-) stellar atmosphere neither of these limiting cases
may be realised, since the fractionation of condensates by
gravitational settling and turbulent mixing compete with each other
and both effects have to be accounted for.  We have therefore
consistently used the latest version of the Settling models, which
model the vertical distribution of condensates by comparing the
dynamical timecales of dust condensation, growth and sedimentation
with vertical mixing by convective overshoot.  The overall shape of
the velocity field in the overshoot region, which is responsible for
supporting cloud layers in the radiative zone, is based on a limited
set of 3D-hydrodynamical simulations \citep{ludwig-MconvII}. Results
of these simulations are generalised to different atmospheric
temperatures and gravities using the velocities predicted by mixing
length theory within the convectively unstable region, and
extrapolating from there with an exponential relation based on the
hydrodynamical models.  The main free input parameter in our models
controls this parametrisation of the overshoot velocity and is chosen
to match the observed changes of dust signatures along the L and T
spectral sequences.

Atmosphere structures for the models were computed solving the
radiative transfer on a medium-resolution wavelength grid of
1\,--\,2\,{\AA} spacing for the optical to near-IR region. From the
converged structures high-resolution spectra with a sampling of
0.05\,{\AA} were produced in one or two additional iterations,
checking for temperature corrections and flux error to ensure that the
final spectral energy output was consistent with the calculated
structure.  We did not fit $T_{\rm eff}$\ and log\,$g$\ to obtain the
best match to the observed spectra, since modelling of high-resolution
spectra is known to produce misleading results for these fundamental
parameters \citep{Schweitzer01,Reiners05}.  Instead we picked the
effective temperature according to the spectral class -- $T_{\rm
  eff}$\ relation of \citet{vrba04,dave04}, which is based on absolute
luminosity and therefore is much less subject to modelling
uncertainties.  The L dwarfs in this study, as isolated field brown
dwarfs, are likely to be fairly old (at least several 10$^8$ years)
and massive, an assumption confirmed by the absence of lithium in
their spectra (cf.\ \ref{sect:alkalis}). Evolutionary models
\citep{evolDust,evolBD} predict gravities in the range $\log g = 5.2
\ldots 5.4$ for such objects.  Therefore a single log\,$g$\,$=$\,5.25
has been used for all models, except for $\epsilon$ Indi\,B, where
$T_{\rm eff}$ and log\,$g$ are much better constrained by the
photometry, parallax and age estimate of \citet{markeIndi}.

\section{Spectral features and comparison to models}

In this section, we discuss the key L and T dwarf features shown in
our spectral atlas in Figs.\,\ref{fig:atlas1}--\ref{fig:atlas10}. We
compare the data to the modeled spectra and try to identify the
improvements necessary to obtain a better fit. We discuss the
molecular features in \S\,\ref{sect:molecules} before we turn to
the atomic lines in \S\,\ref{sect:atoms}.

\subsection{Molecular bands}
\label{sect:molecules}

\subsubsection{Metal Oxides}

TiO and VO are the most important opacity sources in M stars. They
remain strong in the L dwarfs, but get weaker towards cooler
temperature due to condensation into dust species like perovskite
(CaTiO$_3$), solid titanium oxides and vanadium oxides, respectively
\citep[see {e.\,g.}][]{loddersTiV,jwfOpac05}.  Our spectra cover the
TiO bandheads at 7053\,\AA, 7589\,\AA, 7666\,\AA, 8432\,\AA, and
8859\,\AA. All bands are recovered by the models, but the TiO bands
are too strong in the theoretical spectra particularly in the L2 and
L5 dwarfs.  This hints at remaining shortcomings in our treatment of
dust settling (see \S\,\ref{sect:models}), which evidently predicts
too much TiO to be left in the atmosphere at temperatures between 1500
and 2000\,K. VO bands are visible at 7334\,\AA, 7851\,\AA, and
8521\,\AA. In general, VO bands match the data quite well, but a
detailed comparison is hampered buy the mismatch of the TiO bands,
which often affects the same spectral regions.

Another uncertainty that may be responsible for parts of the mismatch
between our data and the model spectra are the (TiO) absorption bands'
oscillator strengths \citep{Allard00, Reiners05}. Laboratory
experiments do not yet provide values at the accuracy required in
order to match the data, and further improvement on the molecular data
is needed.

\subsubsection{Metal Hydrides}

The three metal hydride species CaH, FeH, and CrH are important in the
optical spectra of L dwarfs. CaH at 6750\,\AA\ is visible in the L
dwarfs and is accurately reproduced by the models. The excellent match
between the L0 model and the spectrum shows that the high frequency
patterns visible at this spectral resolution (around 7000\,\AA) is the
structure of the absorption band and not noise. 
These patterns disappear around L5, confirming that CaH is depleted at
later spectral types and is superseded by the K\,\textsc{i} satellite
feature (see \S\,\ref{sect:alkalis}). 

The general structure of the CrH band at 8611\,\AA\ also matches quite
well the structure of the data in all three L dwarfs and also in the T
dwarf.  Unfortunately, this band is embedded in the strong TiO and VO
bands, which the models overpredict, as well as FeH. The calculated
strength of the CrH hence is difficult to judge in the L dwarfs (see
below). 

FeH is observed close to the CrH band at 8692\,\AA\ and extends to the
red all the way to $>1\mu$m, but is increasingly superposed by CrH,
and between 8861\,--\,9334\,\AA\ by TiO. The models compare to the
data as good as for CrH. The prominent Wing-Ford band of FeH is
visible at 9896\,\AA. Structure and intensity of this band are well
fit by the model at the first few {\AA ngstrom}, in early L dwarfs.
However, this band is overestimated at very low temperature as well.
At 9969\,\AA, another CrH band should be visible. It is contained in
the model spectra dominating the absorption redwards of 9969\,\AA\ in
the L5 and T1 spectra. However, \cite{Reiners06a} show that all
absorption features at that wavelength region can be explained by FeH
in M-dwarfs, they see no evidence for CrH absorption. In the models
shown here, CrH produces a strong step in the model spectra for the L5
and T1 objects. Such a feature at 9969\,\AA\ is not observed in our
data, or at least at much less strength, supporting the argument that
the absorption around 1\,$\mu$m in early L dwarfs is entirely due to
FeH, and that CrH is much weaker than predicted.

A similar case can be made for the 8692\,\AA\ band.  While for the
earlier Ls the agreement with the observed band seems reasonably good,
in the T dwarf system it is much too strong in the model. A more
detailed investigation of the molecular line formation in the models
reveals that at higher $T_{\rm eff}$ both FeH and CrH contribute in
similar shares to the absorption all the way to the blue end, with FeH
being mostly responsible for the stronger lines and the structure of
the band. In the 1280\,K model however, settling effects have already
removed significant quantities of the hydrides from the line forming
region. FeH is much stronger affected by this, being depleted already
two pressure scale heights deeper than CrH, which is equivalent to
about one order of magnitude in optical depth.  Thus, in the T1 dwarf
the band is probably dominated by CrH. We therefore conclude that we
see the combined effects of FeH with fairly accurate opacity data, and
CrH absorption that appears overestimated by a significant factor, in
our spectral sequence.  Future modelling efforts for improvement in
this spectral region should therefore focus on better oscillator
strengths for the CrH lines.

\subsubsection{Water}

Strong steam bands are visible at very low temperatures in the
wavelength region redward of 9200\,\AA, where TiO and CrH are also
important (see above). The model spectra provide a rather poor fit to
the data, although the positions of lines are sometimes well
reproduced in regions where no telluric absorption appears (e.g.
around 9250\,\AA).  However, the main contribution to the absorption
in the L models turns out to be due to TiO and CrH, which have both
been shown to be overestimated. The total strength of the remaining
flux matches quite well at large parts of these spectral regions.

\citet{Burgasser03} identified a water band at 9250\,\AA. Our models
and the L dwarf spectra show that this feature is probably the (102)
vibrational level at 10868.88\,cm$^{-1}$
\citep[9200.6\,\AA;][]{Auman67, 2005JMoSp.233...68T}. In the models,
water is not the only opacity source below 9280\,{\AA}, but FeH and
CrH are still strong. This spectral range definitely requires further
attention before it can be used for analysis.

\subsection{Atomic lines}
\label{sect:atoms}

\subsubsection{Alkali lines}
\label{sect:alkalis}

The strongest absorption lines in the spectra of ultra cool stars are
the massively pressure broadened alkali lines, i.e. elements with only
one electron in their outer shell, which can get excited even at very
low temperatures.  All alkali elements have strong lines in the
observed wavelength region. The first alkali element, Li, has a
prominent resonance line at 6708\,\AA. This line is used as an
indicator of brown dwarf status in young objects
\cite[e.g.][]{Basri00}, since evolutionary models agree that stars and
brown dwarfs of more than 0.055\,--\,0.06\,$M_\odot$ will burn a
significant fraction of their primordial lithium content within at
most a few 10$^8$ years \citep{2005AN....326..948Z}.  The brown dwarfs
in our sample, being field objects, can be generally expected to be
old and massive enough to have been depleted of most Li. This is not
included in our model atmospheres, which have been calculated
independently of any assumptions on the evolutionary status. The
theoretical spectra thus show the Li\,\textsc{i} resonance line at
full strength, while comparison with the observations demonstrates
that the actual Li content is severely reduced.  $\epsilon$\,Indi~Ba
is a limiting case in this respect, as the preferred mass estimate of
this brown dwarf, corresponding to a system age of 1.3\,Gyr
\citep{markeIndi}, places it somewhat below the Li burning threshold.
A mass slightly above the limit is still in full agreement with
observed properties and evolutionary models. $\epsilon$\,Indi~Bb
contributes only $\sim$\,1\,\% of the flux at this wavelength, hence
its Li\,\textsc{i} line can not be observed in the combined spectrum.

The Na resonance lines at 5891\,\AA\ and 5897\,\AA\ are not covered by
our spectra. The two subordinate lines at 8194\,\AA\ and 8183\,\AA,
however, are covered. In our setting, they appear just at the red end
of the lower chip of the red arm. Both Na lines are clearly seen in
the L0, the L2, and the L5 dwarf. In the T1 dwarf, they are detected
as narrow absorption spikes. The models predict rather well the shape
of these lines in the L0 and L2 dwarf while their strength is not very
well fit. In the L5 and T1 dwarfs, the observed lines are broader and
shallower than those predicted by the models.

The K\,\textsc{i} resonance lines at 7665\,\AA\ and 7698\,\AA\ govern
the spectral shape of ultra-cool dwarf spectra in the optical
wavelength regime.  These lines have equivalent widths of several
hundred {\AA ngstrom} in T dwarfs and are evolving from narrow lines
in K and M dwarfs to very wide and smooth absorption troughs in late L
and T dwarfs with wings extending more than 1000\,\AA\ to the blue and
especially to the red, as predicted by the line profile calculations
described in \S\,\ref{sect:models}.  The qualitative behaviour of the
K\,\textsc{i} doublet is reproduced well by the model fits, especially
the strong pressure broadening wings in the late L and T dwarfs. Our
new models predict the position of the K-H$_2$ quasi-molecular
satellites at 6900\,--\,7100\,\AA\ (at the position of the O$_2$
B-band). This wide and shallow feature, covering several echelle
orders, may be present in our high-resolution spectra of
$\epsilon$\,Indi~B, but a detailed analysis of this feature goes
beyond the scope of this paper.

The shape of the K lines in the early and mid-L spectra is not that
well matched by the models. In the L0 dwarf, the cores of the observed
K lines are still visible as relatively narrow absorption minima
embedded in wings extending a few tens to hundred \AA. This broader
absorption component becomes saturated already the L2 dwarf spectrum,
eating up the core feature. The models qualitatively predict this
effect, however, they predict the intermediate wings to be much
stronger than actually observed. In the low resolution plot in
Fig.\,\ref{fig:atlas_full}, one can see that this part of the wings is
still overpredicted in the L5 spectrum, while the far wings, that
affect the spectrum all the way to the $Y$ band peak, are quite well
reproduced.  Since the central parts of the lines form in the highest
layers of the atmosphere, especially in the early L dwarfs where
alkali metals are not depleted strongly in the upper regions
\citep{Johnas07}, this is a clear indication that the Settl models do
not yet produce correct densities of the neutral alkali metals in the
uppermost part of the atmosphere.

Absorption lines of Rb at 7800\,\AA\ and 7947\,\AA, and of Cs at
8521\,\AA\ and 8943\,\AA\ appear in all four spectra, even in the T
dwarf spectrum. They become stronger and wider towards lower
temperature. Given the fact that they are embedded in pseudocontinua
that are not always a good fit to the data, the behavior of all Rb and
Cs lines is very well reproduced by the models. A particularly good
example is the Rb\,\textsc{i}~7800\,\AA\ line in the L5 dwarf. This
line very accurately reproduces the data.

\subsubsection{Other Lines}

Hydrogen is observed in emission in some low mass objects, it is an
indicator of a hot chromosphere that is probably magnetically heated.
We find H$\alpha$ in emission in our L0 and L2 spectra, which is
occasionally seen in early type L dwarfs \citep{Schmidt07}.  In the
two cooler objects, no line is detected. This feature does not come
from the photosphere itself, hence it is not contained in the model
spectra.

The next set of lines of the periodic table that can be expected in
spectra of ultra cool dwarfs are the features of neutral earth alkali
elements; these are Be, Mg, Ca, Sr, and Ba. Be, Mg, and Sr do not have
any transitions from low level energy states in the optical wavelength
region, and we are left with transitions of Ca~\textsc{i} and
Ba~\textsc{i}. One Ca~\textsc{i} line is located at 6572\,\AA,
immediately next to H$\alpha$. The line is predicted in the models, it
is only a little weaker than predicted in the data. Two lines of
Ba~\textsc{i} were seen by \cite{Reiners06b} in an L sub-dwarf at
7912\,\AA\ and 8560\,\AA. We do not see these lines in any of our
spectra, which can be explained by the stronger absorption due to the
metal oxides in that region against which the weak Ba lines cannot be
detected \citep{Reiners06b}.

Apart from these lines, we found only a set of low energy Ti\textsc{i}
lines. They are located between 9600 and 9700\,\AA\ and belong to low
energy transition that appears to be visible in such cool atmospheres
\citep[see also][]{Reiners06b}.

The models predict a rather strong $\lambda$6894.5\,\AA\
Sr\,\textsc{i} line, which we do not detect. This might indicate that
the condensation chemistry of this earth alkali metal is also not
fully understood, but the SNR of the observations in this region is
low and telluric absorption by the O$_2$\,B band hampers a detailed
comparison.

\section{Summary}

We have presented high resolution spectra of four ultra-cool dwarfs of
spectral type L and T. The complex chemistry and its feedback to the
structure of such cold atmospheres is a great challenge for theorists
from different areas. Our comparison to the most recent calculations
showed that the general features are understood and that probably most
of the species and their spectral effects are taken into account.
However, in individual features, a good match between data and model
spectra are still not the rule. While calculations of molecular VO and
CaH, and of atomic Cs and Rb are already quite satisfying, features
due to molecular TiO, CrH, water, and atomic Li, Na, and K show large
differences to the observed data.

Particularly in ultra-cool atmospheres, it is necessary to achieve a
very good fit in all important absorbers in order to determine
atmospheric properties, because the chemical compexity of these
atmospheres reacts very sensitively on changes in any of its
constituents. Thus, the full information of spectroscopic features in
ultra-cool dwarfs still cannot be utilized to derive physical
parameters like for example gravity, but we have clearly witnessed
huge progress during the last years. With high quality spectra like
the ones presented in this work, we can continue to improve detailed
models of ultra-cool dwarf atmospheres.

\begin{acknowledgements}
  AR has received research funding from the European Commission's
  Sixth Framework Programme as an Outgoing International Fellow
  (MOIF-CT-2004-002544), and from the DFG as an Emmy Noether Fellow
  (RE 1664/4-1).  We thank the Gesellschaft f{\"u}r Wissenschaftliche
  Datenverarbeitung G{\"o}ttingen for generous allocation of computing
  time used for our calculations.  This work has benefitted from the
  M, L, and T dwarf compendium housed at \texttt{DwarfArchives.org}.
\end{acknowledgements}

\end{document}